\documentclass[11pt,axodraw]{article}
\usepackage{graphicx}
\usepackage{dcolumn}
\usepackage{amsmath}
\usepackage{epsfig}
\usepackage{axodraw}

\input pubboard/babarsym
\input Definitions

\newcommand{\BABARPubYear}    {04}

\newcommand{\BABARConfNumber} {46}
\newcommand{\SLACPubNumber} {10621}
\newcommand{\LANLNumber} {xxxx}

\long\def\inst#1{\par\nobreak\kern 4pt\nobreak
    {\it #1}\par\vskip 10pt plus 3pt minus 3pt}

\begin{document}
{\pagestyle{empty}

\begin{flushright}
\babar-CONF-\BABARPubYear/\BABARConfNumber \\
SLAC-PUB-\SLACPubNumber \\
hep-ex/\LANLNumber \\
July 2004 \\
\end{flushright}

\par\vskip 5cm

\begin{center}
\Large \bf Evidence For 
\boldmath$B^0\to\rho^{0}\KS$ 
\end{center}
\bigskip

\begin{center}
\large The \babar\ Collaboration\\
\mbox{ }\\
\today
\end{center}
\bigskip \bigskip

\centerline{\small{\bf Abstract}} 
\vspace{0.1cm}
\noindent
{\normalsize
We present evidence for the decay $B^0\to\rho^{0}\KS$.
The results are obtained from a data sample of $227\times10^6$ 
$\FourS \to B\Bbar$ decays  collected with the \babar\  detector at the \pep2 asymmetric-energy $B$~Factory at SLAC.
From a maximum-likelihood fit giving a yield of $99\pm19$ events and efficiency estimated from simulation we make a preliminary measurement of the 
branching fraction ${\cal B}(\Bz\to\rho^{0}K^0)=(5.1\pm1.0\pm1.2)\tmsix$ where the first error is statistical and the second systematic. The hypothesis of zero signal in the $\rho^0$ mass region, $600MeV-930MeV$, is excluded at the $6.1\sigma$ level. Allowing a $\Bz\rightarrow{f_0}(600)\KS$ contribution in the fit allows us to exclude the hypothesis of zero $\Bz\rightarrow\rho^0\KS$ at the $3.5\sigma$ level.
}

\vfill
\begin{center}

Submitted to the 32$^{\rm nd}$ International Conference on High-Energy Physics, ICHEP 04,\\
16 August---22 August 2004, Beijing, China

\end{center}

\vspace{1.0cm}
\begin{center}
{\em Stanford Linear Accelerator Center, Stanford University, 
Stanford, CA 94309} \\ \vspace{0.1cm}\hrule\vspace{0.1cm}
Work supported in part by Department of Energy contract DE-AC03-76SF00515.
\end{center}

\newpage

\begin{center}
\small

The \babar\ Collaboration,
\bigskip

%
B.~Aubert,
R.~Barate,
D.~Boutigny,
F.~Couderc,
J.-M.~Gaillard,
A.~Hicheur,
Y.~Karyotakis,
J.~P.~Lees,
V.~Tisserand,
A.~Zghiche
\inst{Laboratoire de Physique des Particules, F-74941 Annecy-le-Vieux, France }
% ...

\end{center}\newpage

\section{Introduction}
\label{sec:Introduction}
In the Standard Model (SM), \CP violation arises from a single
phase in the three-generation Cabibbo-Kobayashi-Maskawa 
quark-mixing matrix~\cite{CKM}. Any measurement indicating additional sources of \CP violation would be evidence for new physics. A number of penguin-dominated decays \cite{NewPhys}(states such as $\phi K^0$,
$\eta^\prime K^0$, $K^+K^-K^0$, $\fzero K^0$ and $\rho^0{K^0}$) offer  
potential for making such an observation: they carry the same weak phase as the
 decay $\Bz\to\jpsi K^0$ \cite{phases}, neglecting CKM-suppressed amplitudes, and therefore the Standard Model predicts their mixing-induced
\CP-violation parameter to be
$-\eta_f\stwob = -\eta_f\times0.74\pm0.05$ \cite{HFAG}. Heavy non-SM 
 particles may  appear in additional penguin diagrams, potentially 
leading to new \CP-violating phases and \CP-violation parameters measurably different from those predicted by the Standard Model.
\par
The two leading diagrams for the channel $\Bz\to\rho^{0}K^0$ are shown in Figure \ref{fig:feyn}. The penguin diagram is expected to dominate.
We present evidence for the decay $B^0\rar\rho^0\KS$ . We take the quasi-two-body (Q2B) approach, restricting ourselves
to the region of the $\pi^+ \pi^- \KS$ Dalitz plot dominated by the $\rho^{0}$ contribution and taking effects due to the interference 
between the $\rho^{0}$ and the other resonances in the Dalitz plot as systematic uncertainties.

\begin{figure}[t]
\begin{center}
\centering{\begin{picture}(500,200)(0,0)


\Text(10,120)[l]{$\bar{b}$}
\ArrowLine(20,120)(60,120)

\Text(80,100)[t]{$W^+$}
\DashLine(60,120)(112,85){5}

\Text(155,105)[r]{$u$}
\ArrowLine(112,85)(142,105)

\Text(155,65)[r]{$\bar{s}$}
\ArrowLine(112,85)(142,65)

\Text(155,120)[r]{$\bar{u}$}
\ArrowLine(60,120)(142,120)

\Text(10,50)[l]{$d$}
\ArrowLine(20,50)(142,50)
\Text(155,50)[r]{$d$}

\Vertex(60,120){2}
\Vertex(112,85){2}

\Text(165,113)[c]{\Large$\}$}
\Text(180,113)[c]{\Large$\rho^0$}

\Text(165,55)[c]{\Large$\}$}
\Text(180,55)[c]{\Large$K_s^0$}

\Text(80,25)[c]{(a)}



\Text(210,120)[l]{$\bar{b}$}
\ArrowLine(220,120)(260,120)

\Text(290,150)[t]{$W^+$}
\DashCArc(285.89,105)(30,30,150){5}

\ArrowArc(285.89,135)(30,270,330)
\ArrowArc(285.89,135)(30,210,270)

\Text(290,85)[bl]{$g$}
\Gluon(285.89,105)(312,85){-4}{5}

\Text(355,105)[r]{$d$}
\ArrowLine(312,85)(342,105)

\Text(355,65)[r]{$\bar{d}$}
\ArrowLine(312,85)(342,65)

\Text(355,120)[r]{$\bar{s}$}
\ArrowLine(312,120)(342,120)

\Text(210,50)[l]{$d$}
\ArrowLine(220,50)(342,50)
\Text(355,50)[r]{$d$}

\Vertex(285.89,105){2}
\Vertex(260,120){2}
\Vertex(312,120){2}
\Vertex(312,85){2}

\Text(365,113)[c]{\Large$\}$}
\Text(380,113)[c]{\Large$K_s^0$}

\Text(365,55)[c]{\Large$\}$}
\Text(380,55)[c]{\Large$\rho^0$}

\Text(286,25)[c]{(b)}

\end{picture}}
\caption{\label{fig:feyn}The two main amplitudes expected to contribute to the decay $B^0\rar\rho^0\KS$ are 
        shown above. These are a colour suppressed tree diagram (a) and a gluonic penguin (b). }
\end{center}
\end{figure}
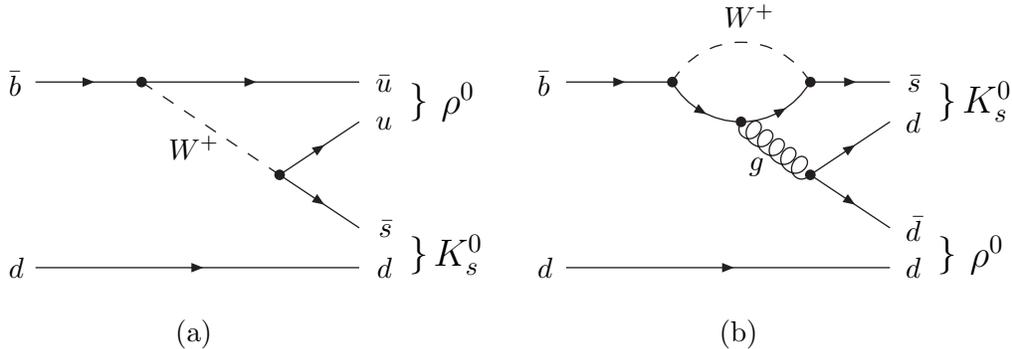

The data we use in this analysis were recorded with the \babar\ 
detector 
at the \pep2 asymmetric-energy $e^+e^-$ 
storage ring at SLAC. The data sample consists of  an integrated 
luminosity of $205\invfb$, corresponding to $(227\pm2)\times10^{6}$
$\B\Bbar$ pairs,  collected at the \FourS resonance
 (``on-resonance'') and $16\invfb$ collected about $40\mev$ below 
the~\FourS (``off-resonance''). 
In Ref.~\cite{bib:babarNim} we describe 
the silicon vertex tracker and drift chamber used for track and vertex
reconstruction, and the Cerenkov detector (DIRC), the electromagnetic 
calorimeter (EMC), and the instrumented flux return (IFR) used for particle 
identification.

\section{The Candidate Selection}
\label{sec:selection}
We reconstruct $B^0\rar \rho^{0} \KS$ candidates ($B^0_{\rm rec}$ in the 
following)  from combinations 
of a $\rho^0$ decaying to $\pi^+\pi^-$ and a $\KS$~decaying to $\pi^+\pi^-$.
For the $\pi^+\pi^-$ pair  from the $\rho^{0}$ candidate, 
we use the combined information  of the tracking system, EMC, and DIRC to 
remove tracks positively identified as electrons, kaons, or protons.  
In addition, we require at least one of the two tracks to have a          
signature in the IFR that is inconsistent with the muon hypothesis.
The mass of the 
$\rho^{0}$ candidate is restricted to the interval $0.6<m(\pi^+\pi^-)<0.93\gevcc$.  To reduce
combinatorial background from low momentum pions, we 
require $|\cos{\theta_{\pi^+}}|<0.95$, where  $\theta_{\pi^+}$ 
is the angle between the directions of the positive pion and the parent $B^0$ in  the $\rho^0$  rest frame.  
The $\KS$ candidate is required to have a mass within $13\mevcc$ of 
the nominal $K^0$ mass \cite{PDG2002} and a decay vertex separated from the $\rho^0$ decay vertex
 by at least three standard deviations. In addition, the cosine 
of the angle between the $\KS$ flight direction and the vector between 
the $\rho^0$ and the $\KS$ decay vertices must be greater than 0.995.

Two kinematic variables are used to discriminate between signal and combinatorial background. 
The first is the difference 
$\de$ between the measured center-of-mass (CM) energy of the $B$~candidate and $\sqrt{s}/2$, where $\sqrt{s}$ is the CM beam energy. The second variable is the 
beam-energy-substituted mass 
$\mes\equiv\sqrt{(s/2+{\mathbf {p}}_i\cdot{\mathbf{p}}_B)^2/E_i^2-{\mathbf {p}}_B^2},$
where the $B$ momentum ${\mathbf {p}}_B$ and the four-momentum of the 
initial $\Upsilon(4S)$ state ($E_i$, ${\mathbf {p}}_i$) are defined in the laboratory 
frame. We require $5.23 < \mes <5.29\gevcc$ and $|\de|<0.15\gev$.

Continuum $e^+e^-\to q\bar{q}$ ($q = u,d,s,c$) events are the dominant 
background.  To enhance discrimination between signal and continuum, we 
use a neural network (NN) to combine five  variables: the 
cosine of the angle between the $B^0_{\rm rec}$ direction 
and the beam axis in the CM,
the cosine of the angle between the thrust axis of the
$B^0_{\rm rec}$  candidate
and the beam axis, the sum of momenta transverse to the direction of flight of 
the $B^0_{\rm rec}$, and the zeroth and second angular moments $L_{0,2}$
of the energy flow about the $B^0_{\rm rec}$ thrust axis.  The moments
are defined by $L_j=\sum_i \mathbf{p}_i \times |\cos{\theta_i}|^j$, 
where $\theta_i$
is the angle with respect to the $B^0_{\rm rec}$ thrust axis of the track
or neutral cluster $i$, and $p_i$ is its momentum.  The sum excludes the tracks that make up the 
$B^0_{\rm rec}$ candidate.
The NN is trained 
with off-resonance data and
simulated signal events. 
The signal efficiency determined from Monte Carlo (MC) 
simulation is $25\%$.  MC simulation shows that  $13.8\%$ of the 
selected signal events are mis-reconstructed,  
mainly due to combinatorial
background from low-momentum background tracks being used to form the $\rho^{0}$ candidate in place of one of the pions it decayed to.
In total, 21000 on-resonance  events pass all selection criteria.

\section{Background from other \boldmath{\B} Decays}
\label{sec:BBackground}

We use high statistics MC-simulated events to study the background from other $B$
decays. The charmless decay modes are grouped into eight classes. The six $B^0$ decay modes to the $\pi^+\pi^-\KS$ 
final state are of particular importance since they have signal-like 
$\de$ and $\mes$ distributions and their decay amplitudes interfere with 
the $\rho^0\KS$ decay amplitude.  Among these modes are $\fzero\KS$, $\ftwo\KS$, $K^{*+}\pi^-$ (including other kaon 
resonances decaying to $\KS\pi^+$), and non-resonant $B^0\rightarrow\pi^+\pi^-\KS$ 
decays.  The inclusive charmless $B^0\rightarrow\pi^+\pi^-\KS$ branching fraction 
$(23\pm3)\tmsix$ together with the available 
exclusive measurements~\cite{HFAG},  are used to infer upper limits on the
contributions of these decays. 
 Selection efficiencies are obtained from MC and used with these branching 
fractions to estimate the expected background.
 The charmed decay  
$B^0\rar D^-\pi^+(D^-\rar\pi^-\KS)$ contributes significantly to the selected data sample despite the veto requiring the invariant mass of both $\KS\pi$ combinations to be more than 40 Mev/$c^2$ from the $D^-$ mass as quoted in \cite{PDG2002}. As a result it is dealt with as an individual component in the fit. Two additional classes account for the remaining neutral and charged $B$ backgrounds, with  $b \to c$ decays being the dominant component in both cases.

\begin{center}
 \begin{table}
  \resizebox{\textwidth}{!}{
\begin{tabular}{|l||c|c|c|}
 \hline
Background Mode & Efficiency ($\%$)& Branching Fraction $(10^{-6})$&  Number of Expected Events \\ 
 \hline
 \hline
$B^0 \rightarrow K_0^*(1680)^- \pi^+$ 	& $0.04\pm0.01$ & $14\pm14$ 	& $2\pm2$ \\ 
$B^0 \rightarrow f_0(980)K_s^0 $	& $0.40\pm0.02$ & $2\pm2$ 	& $2\pm2$ \\ 
$B^0 \rightarrow K_2^*(1430)^+ \pi^-$ 	& $0.20\pm0.02$ & $14\pm14$ 	& $6\pm6$ \\ 
$B^0 \rightarrow f_2(1270) K_s^0$ 	& $1.68\pm0.04$ & $1.7\pm1.7$ 	& $6\pm6$ \\ 
$B^0 \rightarrow K_s^0 \pi^+ \pi^- $ 	& $1.11\pm0.10$  & $2.8\pm1.7$ 	& $7\pm4$ \\ 
$B^0 \rightarrow K^*_0(1430)^+ \pi^-$ 	& $0.13\pm0.01$ & $14\pm14$ 	& $2\pm2$ \\ \hline
$B^0 \rightarrow \eta' K_s^0$ 		& $6.29\pm0.10$ & $6.6\pm0.9$ 	& $94\pm13$ \\ 
$B^0 \rightarrow \rho^0 K^{*0}$ 	& $0.64\pm0.03$ & $7.1\pm7.1$ 	& $10\pm10$ \\ 
$B^0 \rightarrow D^- \pi^+$ 		& $0.34\pm0.02$ & $42\pm6$ 	& $32\pm8$ \\
Charmed $B^0$ 				& 		& 			& $171\pm86$\\
Charmed $B^+$ 				& 		& 			& $106\pm58$\\
\hline
\end{tabular}

}
 \caption{Table of the B background modes included in the Maximum Likelihood fit. The first six modes in this table all decay to the final state $\pi^+\pi^-\KS$ . The number of expected events refers to the full fit region. The error on the number of expected events (used to calculate the systematic error) is taken from the measured branching fraction or is  set to 100\% when the value is not obtained from a direct branching fraction measurement.}
 \label{efficiencyTable}
\end{table}
\end{center}

\section{The Maximum-Likelihood Fit}
\label{sec:themlfit}
We use an unbinned extended-maximum-likelihood fit to extract
the $\rho^{0}\KS$ event yield. In view of a future analysis of time-dependent CP asymmetry, the events in this sample are flavour-tagged as $B^0$ or $\Bzb$ with the method described in \cite{bib:BabarS2b}. Four flavour tagging categories and an ``untagged'' category are defined each having a different expected purity for the signal.
The likelihood function for the $N_\cat$ candidates in flavour tagging category $k$ is
\begin{equation}
\label{eq:pdfsum}
{\cal L}_k = e^{-N^{\prime}_\cat}\!\prod_{i=1}^{N_\cat}
		\bigg\{ N_{S}\epsilon_\cat\left[
                                (1-f^\cat_\textrm{MR}){\cal P}_{i,\cat}^{S-\textrm{CR}} +
                                f^\cat_\textrm{MR}{\cal P}_{i,\cat}^{S-\textrm{MR}}
                              \right]
	+ N_{C,\cat} {\cal P}_{i,\cat}^{C} 
 	+ \sum_{j=1}^{n_B} N_{B,j} \epsilon_{j,\cat}{\cal P}^{\B}_{ij, \cat}
	\bigg\}
\end{equation}
where $N^{\prime}_\cat$ is the sum of the yields of all components, signal and backgrounds,  
tagged in category $\cat$,
 $N_S$ is the number of 
$\rho^{0}\KS$ signal events in the sample, $\epsilon_\cat$ is the 
fraction of signal events tagged in category $\cat$, $f^\cat_\textrm{MR}$
is the fraction of mis-reconstructed signal events in tagging category
$\cat$, $N_{C,\cat}$ 
is the number of continuum background events  that are tagged in 
category~$\cat$, and $N_{B,j}\epsilon_{j,\cat}$ is the number of
$B$-background events of class $j$ that are tagged in category~$\cat$.
The \B-background event yields are fixed in the default fit.
The total likelihood 
${\cal L}$ is the product of the likelihoods for each tagging category.

The probability density functions (PDFs) ${\cal P}_{\cat}^{S-\textrm{CR}}$,  
${\cal P}_{\cat}^{S-\textrm{MR}}$,
${\cal P}_{\cat}^{C}$ and ${\cal P}^{\B}_{j, \cat}$, for correctly reconstructed
signal, mis-reconstructed signal, 
continuum background and $B$-background class $j$ (see Table \ref{efficiencyTable}), respectively,
are each the product of the PDFs of four discriminating variables.
Each signal and background PDF is thus given by:
$  {\cal P}_\cat = 
	{\cal P}(\mes)\cdot 
	{\cal P}(\de) \cdot
 	{\cal P}_\cat(\NN) \cdot 
	{\cal P}(|\cos(\theta_{\pi^+})|)$ 
where \mes, \de, \NN and $|\cos(\theta_{\pi^+})|$ are the variables described in section 2.

The $\mes$, $\de$, \NN, $|\cos{(\theta_{\pi^+})}|$  PDFs 
for signal and $B$ background are taken from the simulation 

There are 15 free parameters in the default fit, including the yields of continuum events in each of the tag categories and the signal yield.
Nine of the free parameters are used to describe the shape of the 
continuum background - third order polynomials to describe the NN shape and 
$|\cos(\theta_{\pi^+})|$, a second order polynomial to describe the $\Delta{E}$
distribution and for $m_{ES}$ a function describing phase space with a single 
free parameter for its slope.

\section{Fit Results}
\label{sec:results}
The maximum likelihood fit results in a signal yield of $99\pm19$, where the signal is the combination of $\Bz\rightarrow\rho^0\KS$ and $\Bz\rightarrow{f_0(600)}\KS$ events. When selection efficiency (25\%), fraction of $K^0$ decaying to $K_s^0$ (50\%), the fraction of $K^0$ decaying to $\pi^+\pi^-$ (69\%) and the initial number of \Bz\Bzb pairs are taken into consideration this leads to: 
\begin{eqnarray*}
{\cal B}(\Bz\to\rho^{0}K^0)=(5.1\pm1.0\pm1.2)\tmsix
\end{eqnarray*}where the first error is statistical and the second systematic. The hypothesis of zero signal in the $\rho^0$ mass region, $600MeV-930MeV$, is excluded at the $6.1\sigma$ level. Allowing as a free parameter the yield of a $\Bz\rightarrow{f_0(600)}\KS$ contribution in the fit, discussed in Section 6, allows us to exclude the hypothesis of zero $\Bz\rightarrow\rho^0\KS$ at the $3.5\sigma$ level.

\begin{figure}
\begin{center}
\scalebox{1.7}{
	\epsfysize5.0cm\epsffile{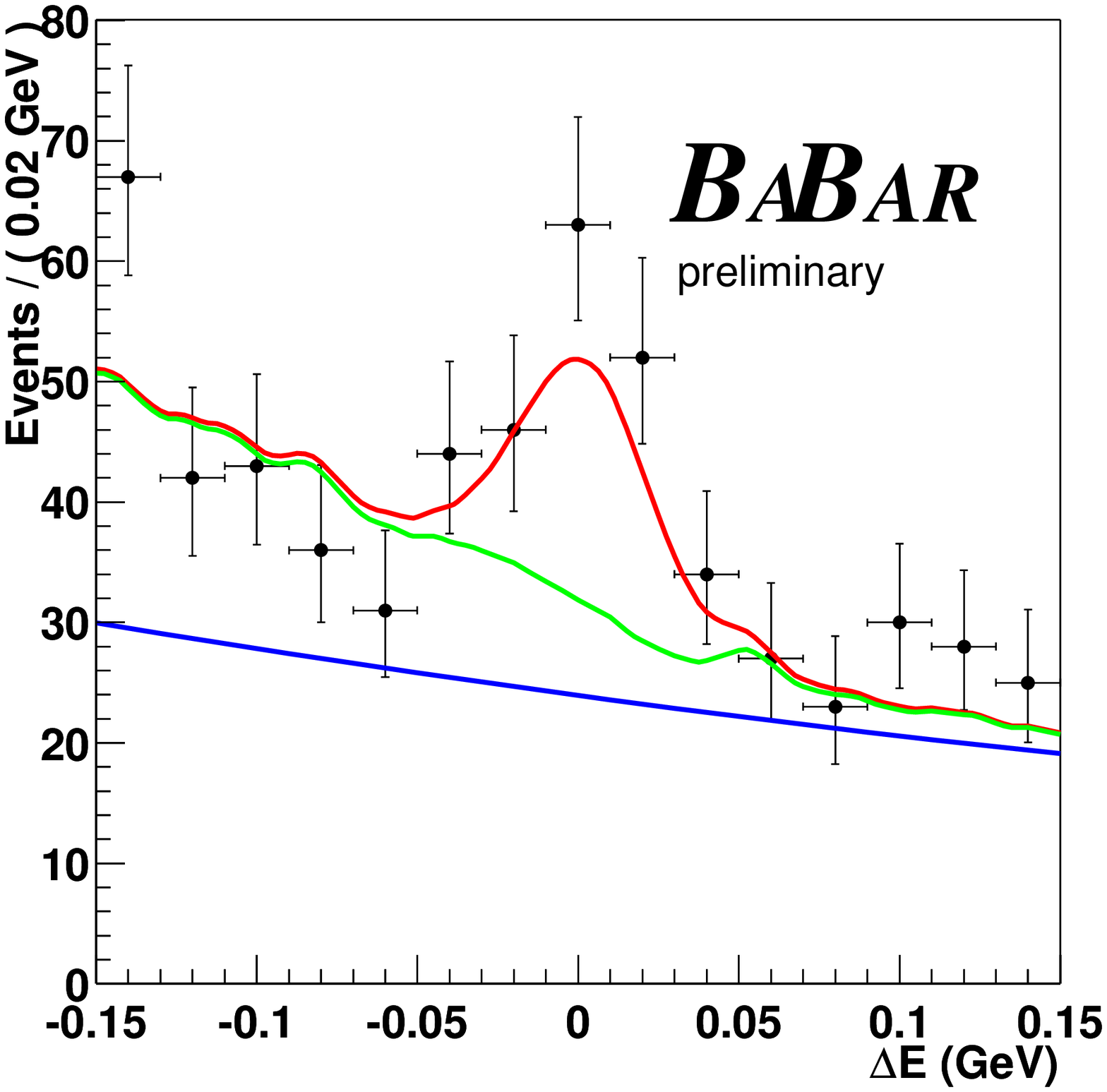}	
	\epsfysize5.0cm\epsffile{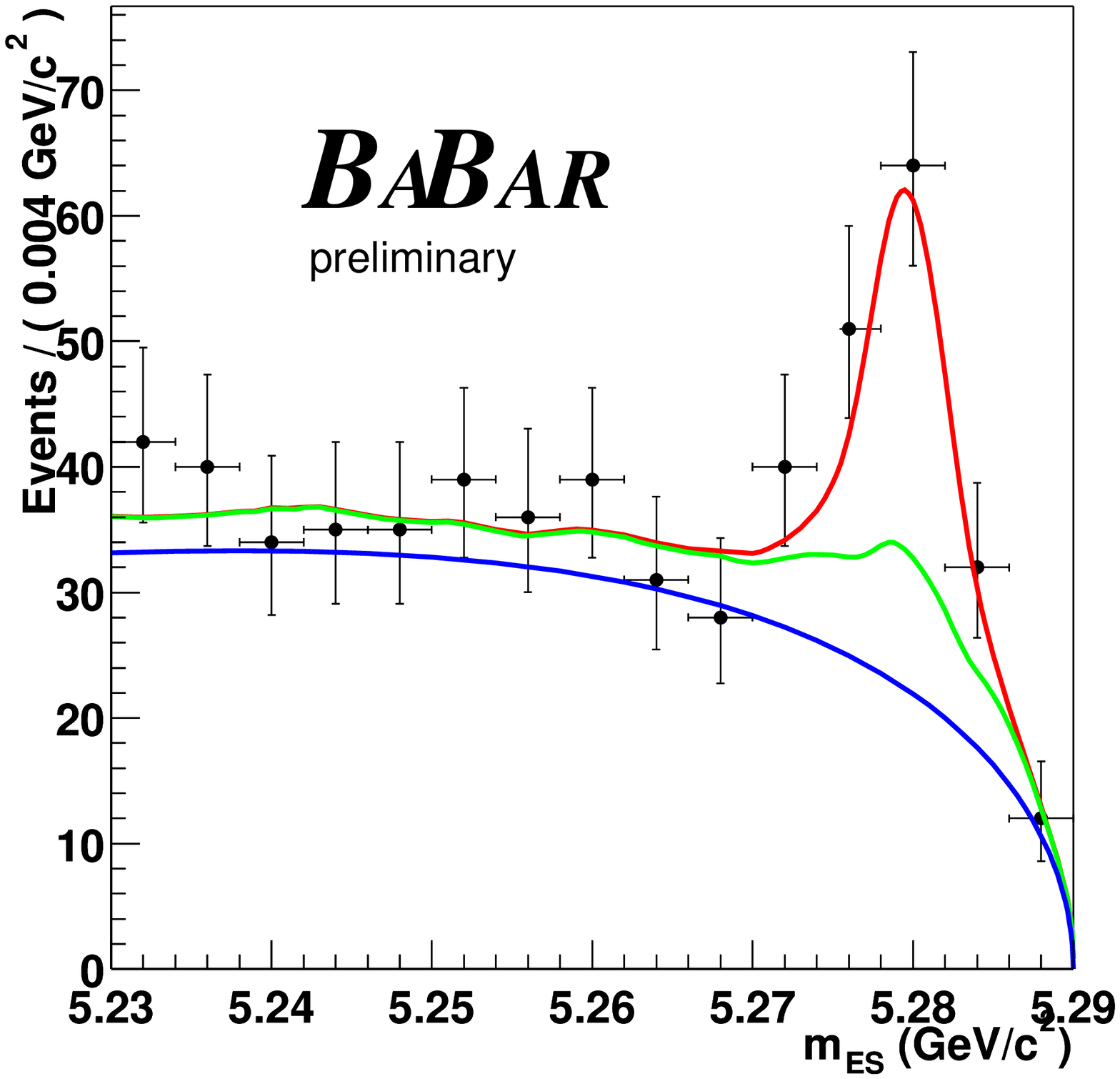}}
\scalebox{1.7}{
 	\epsfysize5.0cm\epsffile{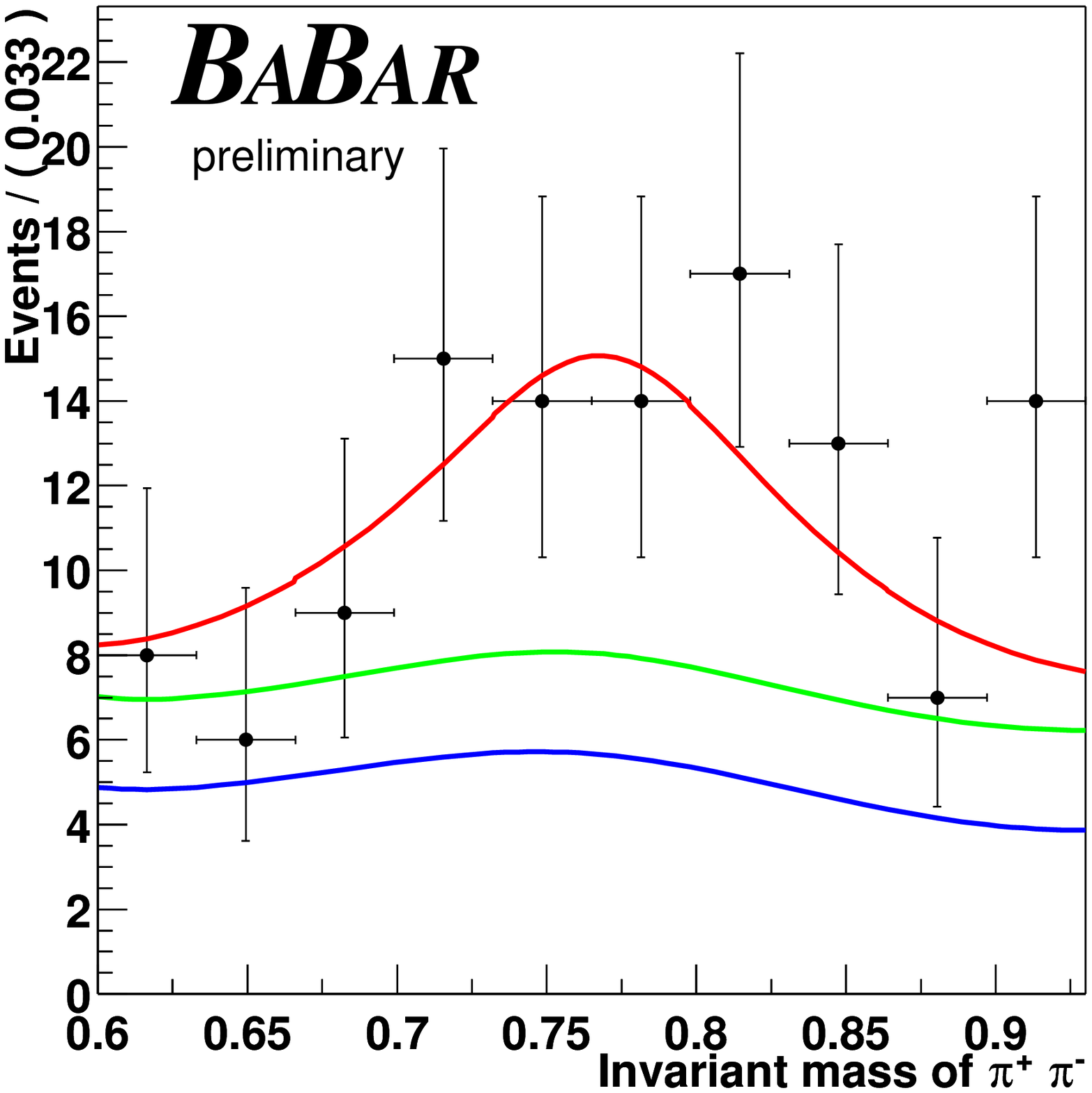}
	\epsfysize5.0cm\epsffile{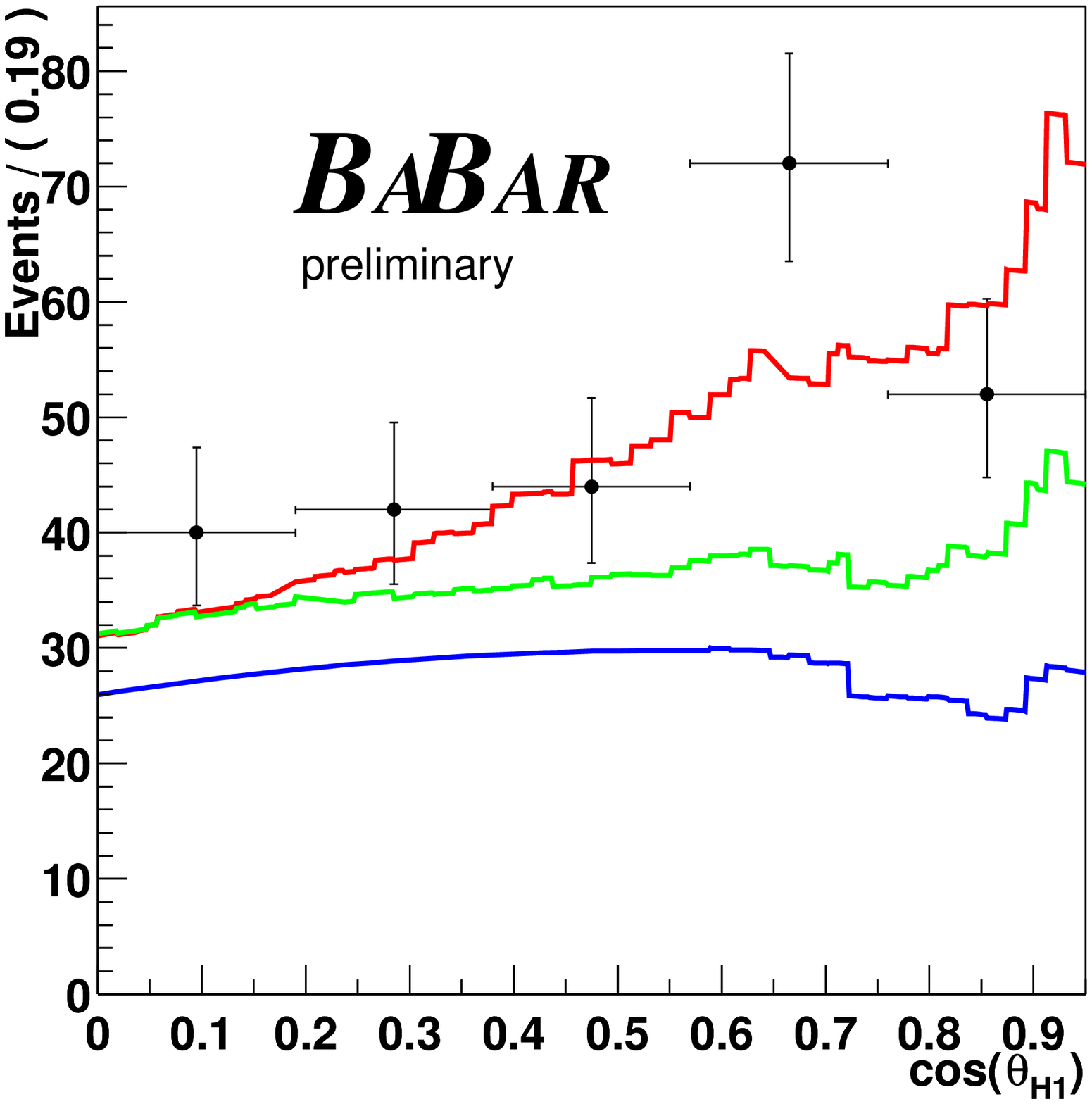}} 
\caption{\label{fig:ProjMesDE}
		Projections of (clockwise from top left) $\de$, $\mes$, 
	$|\cos{\theta(\pi^+)}|$, $m_{\pi^+\pi^-}$ enhanced in $\rho^{0}\KS$ signal through a cut on the ratio of signal to background likelihood using all discriminating variables except the one plotted.	The red (upper) curve represents 
  a projection of the maximum-likelihood fit result. The blue (lower) 
	curve represents the contribution from continuum events, and 
	the green line (middle) indicates the combined contributions from 
	continuum events and $B$ backgrounds. }
\vspace{-0.20in}
\end{center}
\end{figure}

Figure ~\ref{fig:ProjMesDE} shows distributions 
of $\de$, $\mes$, $|\cos{\theta_{\pi^+}}|$ and $m_{\pi^+\pi^-}$,
that are enhanced in signal content by cuts on the signal-to-continuum 
likelihood ratios of the other discriminating variables. 

We performed a number of validation fits with different 
fit configurations to test the stability of the nominal fit;
the results
are consistent within errors.  We made use of simulated experiments where events are generated from the PDFs for signal, continuum and $B$ Backgrounds (toy Monte Carlo). The likelihood of our sample was found to be in good agreement with that obtained from toy Monte Carlo.
\par
Among the additional tests we have performed are: 
\begin{enumerate}
\item Adding $\Bz\to\rho^{0}K^0$ simulated events to data to check that the fitted yield increases with the number of added events. 
\item Fitting for $\Bz\to{f}_0(980)\KS$ and observing a yield consistent with our previous analysis \cite{f0ksprl}.
\item Fits where background yields are allowed to vary: this is attempted with $\Bz\to\eta{'}\KS$, charged and neutral charmed $B$ contributions and $f_2(1270)\KS$. Background yields are consistent with the estimated values used in the default fit, and signal yield does not change by a significant amount.
\end{enumerate}

\section{Systematic Uncertainty}
\label{sec:systematics}

\begin{table}
\begin{center}
\caption{ \label{tab:systematics}
        Summary of systematic uncertainties.}
\small
\begin{tabular}{lc} \hline
& \\[-0.3cm]
Error Source & Error on $BF$\\
\hline
& \\[-0.3cm]
Fitting Procedure           & 3\%   \\
$B$-background              & 9\%   \\
Signal Model                & 4\%   \\
Q2B Approximation           & 4\%   \\
\hline
$f_0(600)\KS$               & 20\% \\
\hline 
Total                       & 23\% \\
\hline \hline

\end{tabular}
\end{center}
\vspace{-0.25in}
\end{table}
The contributions to the systematic error on the signal yield are 
summarized in Table~\ref{tab:systematics}.
To estimate the errors due to the fit procedure, we perform fits on 1000 toy Monte Carlo samples with  the proportions
of  signal, continuum and $B$-background events measured from data. 
A bias of $(3\pm1)\%$ is observed in these fits and the sum in quadrature of the bias and its error is assigned as a systematic uncertainty. We also perform similar fit tests using fully simulated signal and $B$-background events added in the correct proportions to toy Monte Carlo continuum events. 
We use this technique to pick up any biases that may escape the conventional toy test: we observe no additional statistically significant bias and hence do not add an additional systematic uncertainty.
\par
For each class of $B$ Background, the expected event yields are varied 
according to the uncertainties in the measured or estimated branching fractions, and the change in signal yield taken as a systematic.
\par
The uncertainties due to the extraction of the signal PDFs from simulation are obtained from a control 
sample of fully reconstructed $B^{0} \rightarrow D^{-}(\to\KS\pi^-) \pi^{+}$
decays. We assume that differences in PDFs between data and Monte Carlo for this mode
imply an equivalent data/Monte Carlo discrepancy in $B^{0} \rightarrow\rho^0\KS$ and use them to estimate a 4\% systematic error.

The systematic error introduced through the use of the quasi-two-body 
approximation, neglecting the interference effects between the $\rho^{0}$ 
and the other 
resonances  in the Dalitz plot, is estimated from simulation.  We use a toy Monte Carlo technique, allowing all relative strong phases to take random values in each experiment. We take the RMS change in the signal yield as the systematic uncertainty.  
All $\Bz$ decays to $\pi^+\pi^-\KS$ expected to provide events that pass selection cuts are included in the study, including $f_0(980)$, $f_2(1270)$, 
and the $K^{*+}\pi^-$ and higher kaon states. 
In addition, a non-resonant $\Bz\to\pi^+\pi^-\KS$ component is included in the study.  
The proportion of each contribution is estimated 
using known exclusive measurements and the inclusive $\Bz\to\pi^+\pi^-\KS$ rate.
\par  
An important systematic effect is the possible presence of $B^{0}\rightarrow{f_0(600)}\KS$, where $f_0(600)$ is used to denote a broad scalar contribution of ill defined width that may lie beneath the $\rho^0$. To determine an upper limit on the size of such an effect we perform a fit with a $B^{0}\rightarrow{f_0(600)}\KS$ contributing a yield allowed to float freely (since we do not include $\pi\pi$ mass in this fit we make no assumption about mass distribution). We fit a yield of $31\pm27$ $B^{0}\rightarrow{f_0(600)}\KS$ events.
We back this up by performing fits where helicity is not used as a discriminating variable in the full helicity range and in the $|cos(\theta_{\pi^+})|<0.5$ regions. From the fit to the full range and the efficiency in the low helicity range (estimated from simulation) we expect 22 events in our data sample and observe $36\pm12$. \par
We take the difference between the default fit and a fit assuming a $f_0(600)\KS$ contribution estimated from this second study as an estimate of this systematic effect. This leads to a $20\%$ systematic uncertainty in the signal yield.

\section{Summary}
In summary, we have presented evidence for $\Bz\to\rho^{0}Ks$. From
a maximum-likelihood fit of the signal yield and efficiency estimated from Monte Carlo we measure the branching fraction ${\cal B}(\Bz\to\rho^{0}K^0)=(5.1\pm1.0\pm1.2)\tmsix$ where the first error is statistical and the second systematic. 
 The hypothesis of zero signal in the $\rho^0$ mass region, 600MeV-930MeV, is excluded at the $6.1\sigma$ level. Floating the yield of a $\Bz\rightarrow{f_0(600)}\KS$ contribution in the fit, discussed in Section 6, allows us to exclude the hypothesis of zero $\Bz\rightarrow\rho\KS$ at the $3.5\sigma$ level.

\section{Acknowledgments}
\label{sec:Acknowledgments}
We are grateful for the 
extraordinary contributions of our \pep2\ colleagues in
achieving the excellent luminosity and machine conditions
that have made this work possible.
The success of this project also relies critically on the 
expertise and dedication of the computing organizations that 
support \babar.
The collaborating institutions wish to thank 
SLAC for its support and the kind hospitality extended to them. 
This work is supported by the
US Department of Energy
and National Science Foundation, the
Natural Sciences and Engineering Research Council (Canada),
Institute of High Energy Physics (China), the
Commissariat \`a l'Energie Atomique and
Institut National de Physique Nucl\'eaire et de Physique des Particules
(France), the
Bundesministerium f\"ur Bildung und Forschung and
Deutsche Forschungsgemeinschaft
(Germany), the
Istituto Nazionale di Fisica Nucleare (Italy),
the Foundation for Fundamental Research on Matter (The Netherlands),
the Research Council of Norway, the
Ministry of Science and Technology of the Russian Federation, and the
Particle Physics and Astronomy Research Council (United Kingdom). 
Individuals have received support from 
CONACyT (Mexico),
the A. P. Sloan Foundation, 
the Research Corporation,
and the Alexander von Humboldt Foundation.

\end{document}